\documentclass[apj,revtex4]{emulateapj}
\usepackage{graphicx}
\usepackage{natbib}
\shorttitle{SN 2014J}
\shortauthors{Sand et al.}

\newcommand{\bmax}{$t(B_{\rm max})$}

\begin{document} 

\title{Post-maximum near infrared spectra of SN 2014J: A search for interaction signatures~$\!$\altaffilmark{*}}

\author{D. J. Sand\altaffilmark{1}, E.Y. Hsiao\altaffilmark{2}, D.P.K. Banerjee \altaffilmark{3}, G.H. Marion\altaffilmark{4}, T. R. Diamond\altaffilmark{5}, V. Joshi \altaffilmark{3},  J.T. Parrent \altaffilmark{6}, M. M. Phillips\altaffilmark{7}, M. D. Stritzinger\altaffilmark{8}, V. Venkataraman \altaffilmark{3} 
}
%
\begin{abstract}
We present near infrared (NIR) spectroscopic and photometric observations of the nearby Type Ia SN 2014J.  
The seventeen NIR spectra span epochs from +15.3 to +92.5 days after $B$-band maximum light, while the $JHK_s$ photometry include epochs from $-$10 to +71 days.  This data is used to constrain the progenitor system of SN 2014J utilizing the Pa$\beta$ line, following recent suggestions that this phase period and the NIR in particular are excellent for constraining the amount of swept up hydrogen-rich material associated with a non-degenerate companion star.  We find no evidence for Pa$\beta$ emission lines in our post-maximum spectra, with a rough hydrogen mass limit of $\lesssim$0.1 $M_{\odot}$, which is consistent with previous limits in SN~2014J from late-time optical spectra of the H$\alpha$ line.  Nonetheless, the growing dataset of high-quality NIR spectra holds the promise of very useful hydrogen constraints.
\end{abstract}


\altaffiltext{*}{Based on observations obtained at the Gemini Observatory under program GN-2014A-Q-8 (PI: Sand).  Gemini is operated by the Association of Universities for Research in Astronomy, Inc., under a cooperative agreement with the NSF on behalf of the Gemini partnership: the National Science Foundation (United States), the National Research Council (Canada), CONICYT (Chile), Ministerio de Ciencia, Tecnolog\'{i}a e Innovaci\'{o}n Productiva (Argentina), and Minist\'{e}rio da Ci\^{e}ncia, Tecnologia e Inova\c{c}\~{a}o (Brazil).}
\altaffiltext{1}{Texas Tech University, Physics Department, Box 41051, Lubbock, TX 79409-1051, USA; david.sand@ttu.edu}
\altaffiltext{2}{Florida State University, Department of Physics, Tallahassee, FL, 32306 USA}
\altaffiltext{3}{Astronomy and Astrophysics Division, Physical Research Laboratory, Navrangapura, Ahmedabad - 380009, Gujarat, India}
\altaffiltext{4}{University of Texas at Austin, 1 University Station C1400, Austin, TX, 78712-0259, USA}
\altaffiltext{5}{NASA Postdoctoral Program Fellow, Goddard Space Flight Center, Greenbelt, MD 20771, USA}
\altaffiltext{6}{Harvard-Smithsonian Center for Astrophysics, 60 Garden St., Cambridge, MA 02138, USA}
\altaffiltext{7}{Carnegie Observatories, Las Campanas Observatory, Colina El Pino, Casilla 601, Chile}
\altaffiltext{8}{Department of Physics and Astronomy, Aarhus University, Ny Munkegade 120, 8000, Aarhus C, Denmark}

\section{Introduction}
\label{sec:intro}

Type Ia supernovae (SNe Ia) result from the thermonuclear explosion of carbon-oxygen white dwarfs \citep{Hoyle60}, and are important cosmological distance indicators.  Despite their intense study, the explosion mechanisms and progenitor scenarios for these events are still not well understood \citep[][]{Maoz14}.  There are two general progenitor scenarios which explain the mechanism by which the white dwarf accretes the necessary mass.  The single-degenerate (SD) scenario posits a non-degenerate companion star 
to the white dwarf \citep{Whelan73}, while the double-degenerate (DD) scenario has two white dwarfs in the pre-supernova system \citep{Iben84,Webbink84}. A third variant possibility, the double detonation explosion mechanism, results from the initial detonation of He on the white dwarf surface, which subsequently triggers carbon detonation in the core \citep[e.g.][]{Shen14}. The initial He detonation comes from accretion from either a degenerate or non-degenerate companion star.



The identification of the progenitor systems of SNe Ia has eluded a clear answer even though many observational techniques have been developed to search for relevant clues \citep[see ][for a concise summary]{Chomiuk15}.
One indirect method for identifying the SD scenario has to do with the fact that part of the hydrogen-rich envelope of the companion star is expected to become unbound via interactions with the SN ejecta, embedded in the inner, low-velocity material \citep{Marietta00}.
This unbound hydrogen material has long been searched for -- via the H$\alpha$ line -- in late-time spectra \citep[e.g.][]{Mattila05,Leonard07,Lundqvist13,Shappee13,Lundqvist15,Graham15,Maguire15}. One tentative detection has been reported \citep[in SN 2013ct;][]{Maguire15}, but only $\sim$15 SNe have strong limits.  This suggests that the SD scenario is sub-dominant, but given that physical limits are reliant on imperfect models \citep[see discussion in][]{Maguire15}, it is essential to investigate alternative proposed hydrogen signatures.



A novel avenue to search for signs of interaction with a companion star was recently presented by \citet[][hereafter M14]{Maeda14}.  Building off the work of \citet{Kasen04}, M14 utilized hydrodynamic simulations of the impact between the SN ejecta and a non-degenerate companion star, along with detailed radiative transfer simulations \citep{Kutsuna15}, to investigate both broadband light curve and hydrogen emission signatures in the maximum and post-maximum period  (up to about 2 months after explosion) in both the optical and NIR.  
M14 suggested that the H$\alpha$ line would be difficult to detect in this time frame, but the expected Pa$\beta$ emission line signature would be increasingly visible in the $\sim$1$-$2 months after maximum light.  
There have been searches for Pa$\beta$ emission in late time spectra \citep[+200 days or more after $B$-band maximum;][]{Lundqvist13,Maguire15}, but not at the epochs suggested by M14.

Intrigued by the prospect of constraining the SN Ia progenitor with post-maximum NIR observations, we present a sequence of high signal-to-noise NIR spectra of SN 2014J, hosted by M82 \citep{Goobar14,Marion15} and 
provide the first constraints on a nondegenerate companion star utilizing the Pa$\beta$ line in the post-maximum period.

We adopt the light curve parameters of SN~2014J determined by \citet{Marion15}, with the time of $B$-band maximum at 2014 February 1.74 UT (MJD= 56689.74); our data is presented with respect to this date throughout this work. We also adopt the light curve decline rate from \citet{Marion15},  $\Delta$$m_{15}$$(B)$ = 1.12 mag, and 
a distance modulus of $\mu$ =
27.64$\pm$0.1 mag 
to M82 based on the average
of the two tip of the red giant branch distance measurements
presented in \citet{Dalcanton09}.  

\section{Near Infrared Photometry} \label{phot}

\begin{figure*}
\begin{center}
\mbox{ \epsfysize=10.0cm \epsfbox{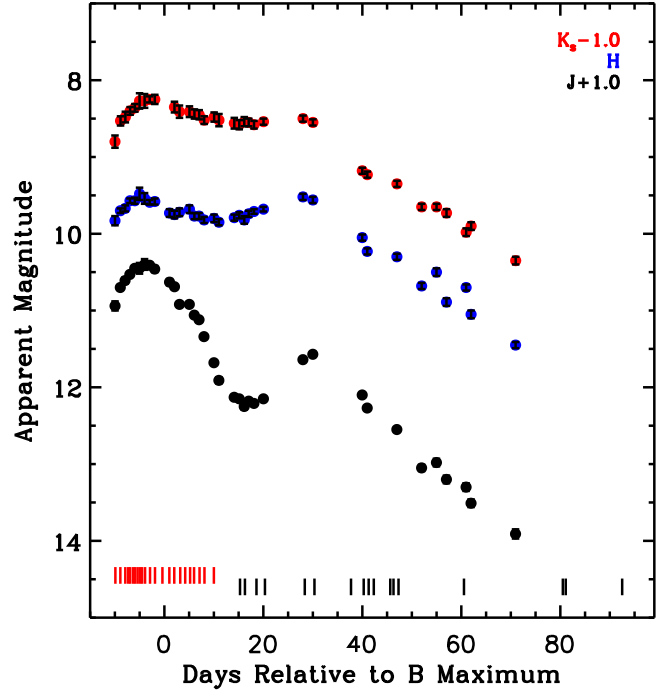}} 
\caption{NIR photometry of SN~2014J from $-$10 to +71d.    
Red vertical lines indicate NIR spectroscopic epochs presented by \citet{Marion15}, while black vertical lines are new NIR spectroscopic epochs.  Those data points after +41d are newly presented in this work. \label{fig:LC}}
\end{center}
\end{figure*}

NIR photometry of SN~2014J was obtained from Mt. Abu Infrared Observatory \citep{Banerjee12} in the $J$, $H$ and $K_s$ bands (using MKO standard filters) utilizing the Near-Infrared Camera/Spectrograph (NICS) on the 1.2m telescope.  
The imaging data was reduced in a standard way, and aperture photometry of the sky-subtracted frames was done with {\sc IRAF}.  The magnitude of SN 2014J was determined differentially, utilizing the nearby Two Micron All Sky Survey star J09553494+6938552 as a reference.  As no difference imaging was performed,  we placed artificial stars into our data in regions with a similar background as that of SN~2014J to assess any potential systematic uncertainties.  This exercise suggests that the $J$ band magnitudes after +54d may suffer systematics of order $\sim$0.1 mag, while those in the $H$ and $K_s$ bands are of order $\sim$0.04 mag. This NIR photometry is a satisfactory match to that presented in \citet{Foley14} around maximum light but there are systematic differences at later epochs, although the Mt. Abu photometry has been shown to be a good match to the NIR light curve of SN~2011fe up to +41d, which is most relevant to our later search for Pa$\beta$ emission (see \S~\ref{sec:search}).  We present our NIR photometry in Figure~\ref{fig:LC} and Table~\ref{table:nirphot}. No correction was made for either Milky Way extinction or the inferred extinction due to M82, as the NIR light curves appear to be nearly extinction-free based on a comparison with the SN 2011fe light curve \citep{Amanullah14}.
Portions of the Mt. Abu photometric NIR light curve have been presented elsewhere \citep{Goobar14,Amanullah14,Marion15}, but the current dataset spans from $-$10d through +71d, extending the lightcurve by $\sim$30 days.  

\section{Near Infrared Spectroscopy}
\label{spec}

\begin{figure*}
\begin{center}
\mbox{ \epsfysize=12.0cm \epsfbox{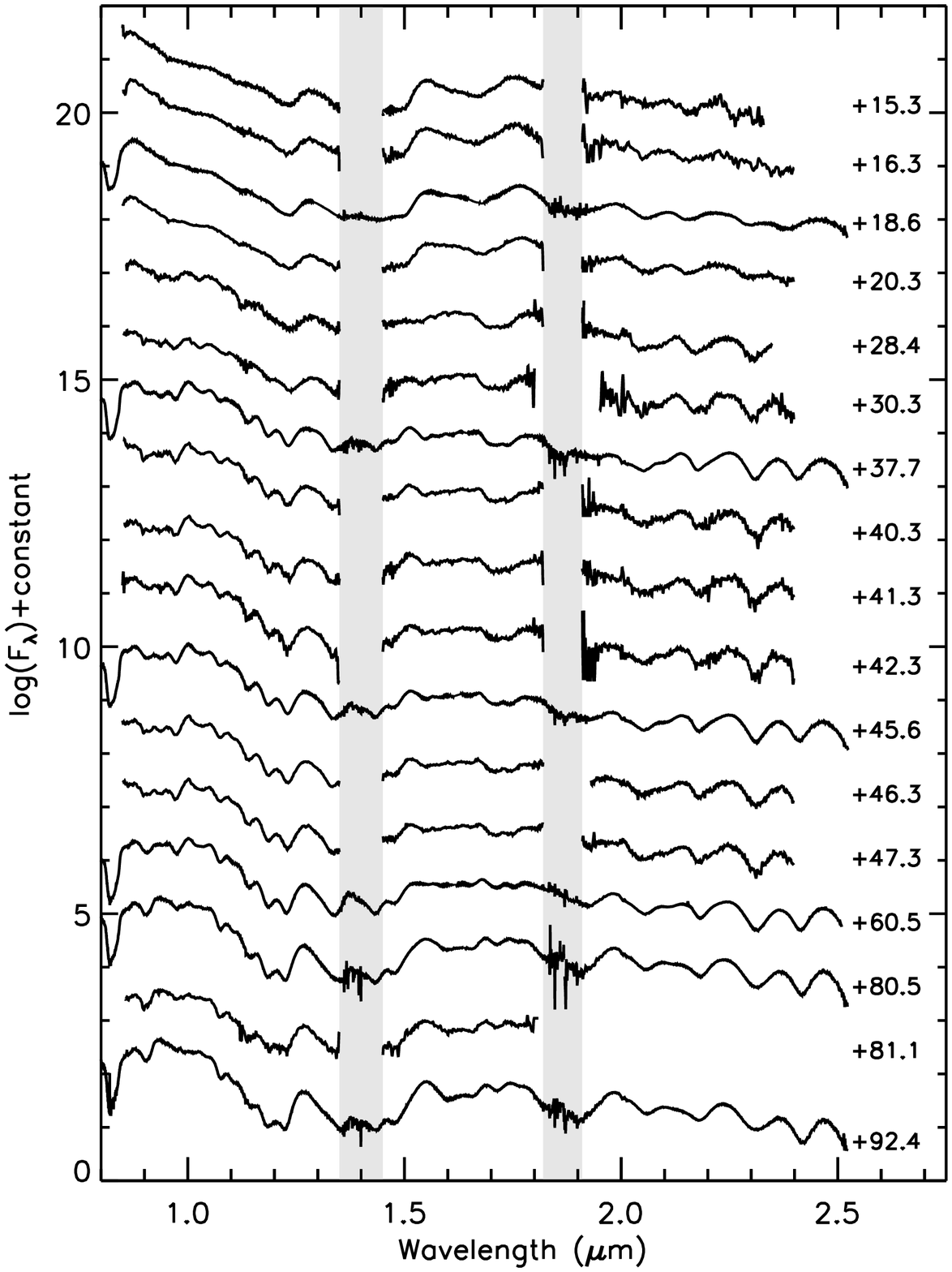}} 
\mbox{ \epsfysize=12.0cm \epsfbox{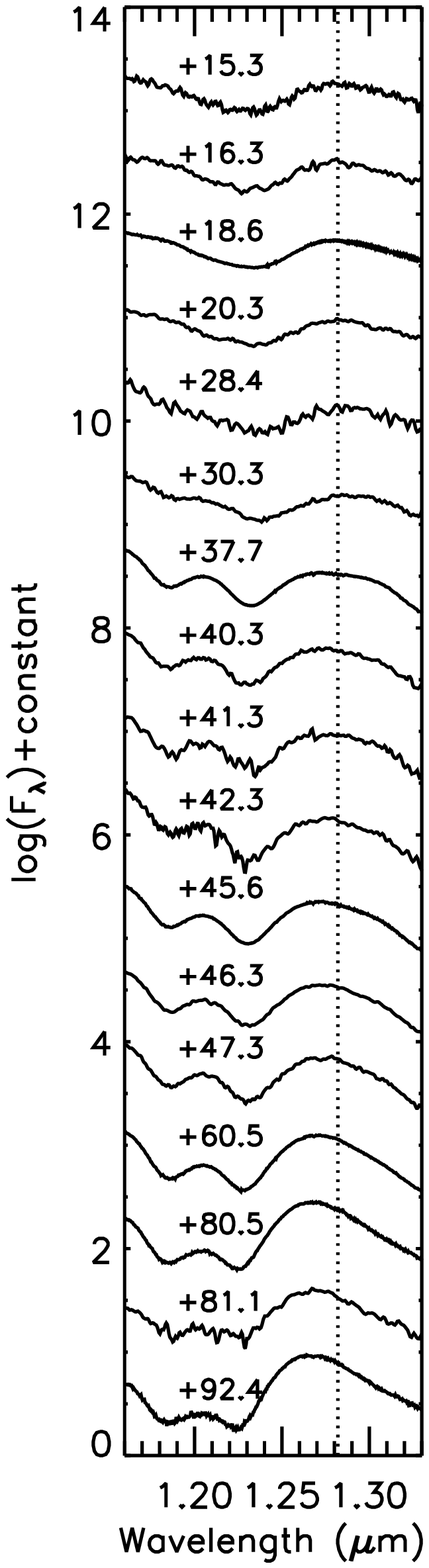}} 
\caption{{\bf Left:} NIR spectra of SN 2014J spanning +15.28d to +92.44d (see Table~\ref{table:spectlog}).  The gray vertical bands mark the regions with strong telluric absorption.  
{\bf Right:} A zoom in on the Pa$\beta$ region, with the rest wavelength marked by the light dotted line.  There are no signatures of Pa$\beta$ emission. 
\label{fig:specs}}
\end{center}
\end{figure*}

We carried out an intensive NIR spectroscopic campaign on SN 2014J, the first results of which were reported elsewhere \citep[spanning $-$10d to +10d;][]{Foley14,Marion15}.  Here we present our remaining post-maximum NIR spectra.  The data include eleven spectra from the Mt. Abu Infrared Observatory and six spectra from the Gemini Near-Infrared Spectrograph (GNIRS) at Gemini North Observatory \citep{GNIRS}. These spectra span from +15.3d to +92.5d, and are shown in Figure~\ref{fig:specs}.  A log of the observations is presented in Table~\ref{table:spectlog}.  All of the spectra are available on WISeREP\footnote{http://wiserep.weizmann.ac.il} \citep{Yaron12}.

The Mt. Abu Infrared Observatory NIR spectra were taken with the Near-Infrared Camera/Spectrograph on the 1.2m telecope, which is equipped with a 1024$\times$1024 HgCdTe Hawaii array.  For the full wavelength coverage of 0.85 to 2.4 $\mu$m, three grating settings were used, with a final resolution of $R$=1000.  All spectra were recorded with the target dithered between two positions along the slit for effective sky subtraction and were reduced in a standard way using IRAF tasks \citep[see][for details]{Das08}.  Observations of an A-type star were used to correct for the effects of telluric absorption.


The GNIRS spectra were taken in cross-dispersed mode, using the 32 l mm$^{-1}$ grating and the 0\farcs675 slit, giving  continuous wavelength coverage from 0.8 to 2.5 $\mu$m and $R$$\sim$1000.  
The observational strategy again consisted of the classical ABBA technique, with the slit positioned along the parallactic angle.  Data reduction was done with the {\sc XDGNIRS} PyRAF-based pipeline provided by Gemini Observatory, which flattens the data, subtracts the sky in the AB pairs, stacks the 2D data and extracts 1D distortion-corrected and wavelength-calibrated spectra.  Corrections for telluric absorption, and simultaneous flux calibration, were accomplished using the IDL software package {\sc xtellcor} and observations of A0V telluric standards using the methodology of \citet{Vacca03}.  

Since the spectra were taken in a variety of photometric conditions and different slit widths, we rescaled each to match the $J$-band magnitude at each epoch (note the Pa$\beta$ line lies in the $J$-band), interpolating the light curve presented in \S~\ref{phot}.  



\section{NIR Companion Signature Search}\label{sec:search}

\subsection{Background}

M14 has highlighted that the Pa$\beta$ line can be used to identify unbound hydrogen rich material  from a companion star in a SN Ia, and provides a stronger/clearer signal than the H$\alpha$ line in the post-maximum time frame.
If the modeling of M14 is correct, the Pa$\beta$ line should be easily observable, growing stronger after maximum light through $\sim$1$-$2 months post-maximum (the time period studied by M14), across a range of viewing angles between the observer, companion star and supernova.  Thus, studying the Pa$\beta$ line at these epochs may provide an efficient path to obtaining progenitor constraints on a statistically significant sample of SNe Ia.  This time period corresponds with the rise and decline of the secondary light curve maximum observed in the NIR \citep[see Figure~\ref{fig:LC} and Figure 18 in][]{Folatelli10}.  

Some observational details of the expected signal in the Pa$\beta$ line can be gleaned from M14.
The model that was focused on (termed `RGa') consisted of a 1 $M_{\odot}$ red giant companion star with a radius of 7$\times$10$^{12}$ cm and a separation from the white dwarf of 2$\times$10$^{13}$ cm at the time of explosion.  This red giant has a 0.4 $M_{\odot}$ He core mass and a 0.6 $M_{\odot}$, convective H-rich envelope.  The companion radius was determined so that it filled its Roche lobe at the time of explosion.  In this model (and the other models presented in M14), $\sim$0.4 $M_{\odot}$ of  H-rich material becomes unbound by the SN interaction, although this is probably an overestimate in their main sequence model due to resolution effects.  The reader is referred to M14 for further details about the hydrodynamic and radiative transfer methodology used.

The post maximum Pa$\beta$ feature was particularly conspicuous in the M14 model at post-maximum epochs, more so than H$\alpha$, both because of the lower optical depths in the NIR and optical contamination from silicon near the putative H$\alpha$ emission.  Comparisons between the models with and without unbound hydrogen yielded flux enhancements of up to $\sim$50$-$60\% if the viewer were aligned with the non-degenerate companion star and supernova at  +25d to +40d.

 To show the observational feasibility of Pa$\beta$ signatures, M14 used archival data of SN~1999ee at +30d and +40d  and compared it with SN~2005cf spectra at similar epochs -- a direct comparison of the spectra is a good measure of the relative amount of hydrogen in the two events, and we will utilize this fact in our SN~2014J search below.  M14 also implanted Pa$\beta$ emission drawn from their hydrodynamic and radiative transfer models (for the red giant companion scenario, RGa) directly into their SN~1999ee data, demonstrating the strength of the signal.   Their rough detection limits corresponded to a hydrogen mass of $\lesssim$0.1 $M_{\odot}$ if viewed from the companion star side and $\lesssim$0.2 $M_{\odot}$ if viewed from the opposite direction.

\subsection{SN 2014J Search}

We now mimic the approach of M14 to constrain the amount of unbound hydrogen material in SN~2014J by both direct comparison with NIR spectra of other SNe Ia at similar epochs and by implanting Pa$\beta$ emission features into our data.  


First, we have visually inspected the spectral region around the Pa$\beta$ line for our entire NIR data set, as can be seen in the right panel of Figure~\ref{fig:specs}.   The dominant spectral feature in this wavelength range is a broad bump from roughly 1.25 to 1.30 $\mu$m which persists throughout our NIR spectral sequence from +15.3d to +92.4d.~~
This feature was identified as \ion{Fe}{2} in the modeling of SN~2005cf done by \citet{Gall12}.~~
According to M14's calculation, excess Pa$\beta$ emission can be visible by eye, especially at epochs around +40d, where Pa$\beta$ is visible at all viewing angles 
-- at least for the optimistic scenario of $\sim$0.3 $M_{\odot}$ of unbound hydrogen material.     
No Pa$\beta$ emission lines are visible in our data set.

Given this lack of visible emission, we turn to direct comparisons with other data sets.  To compare the region around Pa$\beta$ with other well-studied SNe Ia, we have searched the literature for NIR spectra taken at similar epochs to our Gemini spectra of SN~2014J at +17.7d, 37.7d and 45.6d.  Spectra from SNe with a S/N $\gtrsim$25 per unit wavelength and within $\sim$4 days of our GNIRS spectra were considered.  Several fast declining SNe (with $\Delta$$m_{15}$$(B)$$>$1.5) do have published NIR spectra at the appropriate epochs, but they show systematically different velocities from SN~2014J and the others presented here, and so we also do not include them in our current comparison.  In future work we will need to confirm and understand any systematic spectral trends in the Pa$\beta$ region in order to perform a statistical search.

Relatively few spectra match all of these criteria, but we utilize data from SN 2011fe \citep{Hsiao13}, SN 1998bu \citep{Hernandez00,Hamuy02}, SN 2004da \citep{Marion09}, SN 1999ee \citep{Hamuy02}, and SN 2003du \citep{Stanishev07}.  
We also include unpublished Gemini/GNIRS and Magellan/FIRE spectra of ASASSN-14lp at +14.4d, +19.6d, +35.3d and +42.4d, which will be presented in a future contribution (Sand et al., in preparation).  These ASASSN-14lp spectra in particular are of similar quality to the SN~2014J GNIRS spectra; an initial analysis of the optical properties of ASASSN-14lp has been presented in \citet{Shappee15}.

\begin{figure*}
\begin{center}
\mbox{ \epsfysize=5.3cm \epsfbox{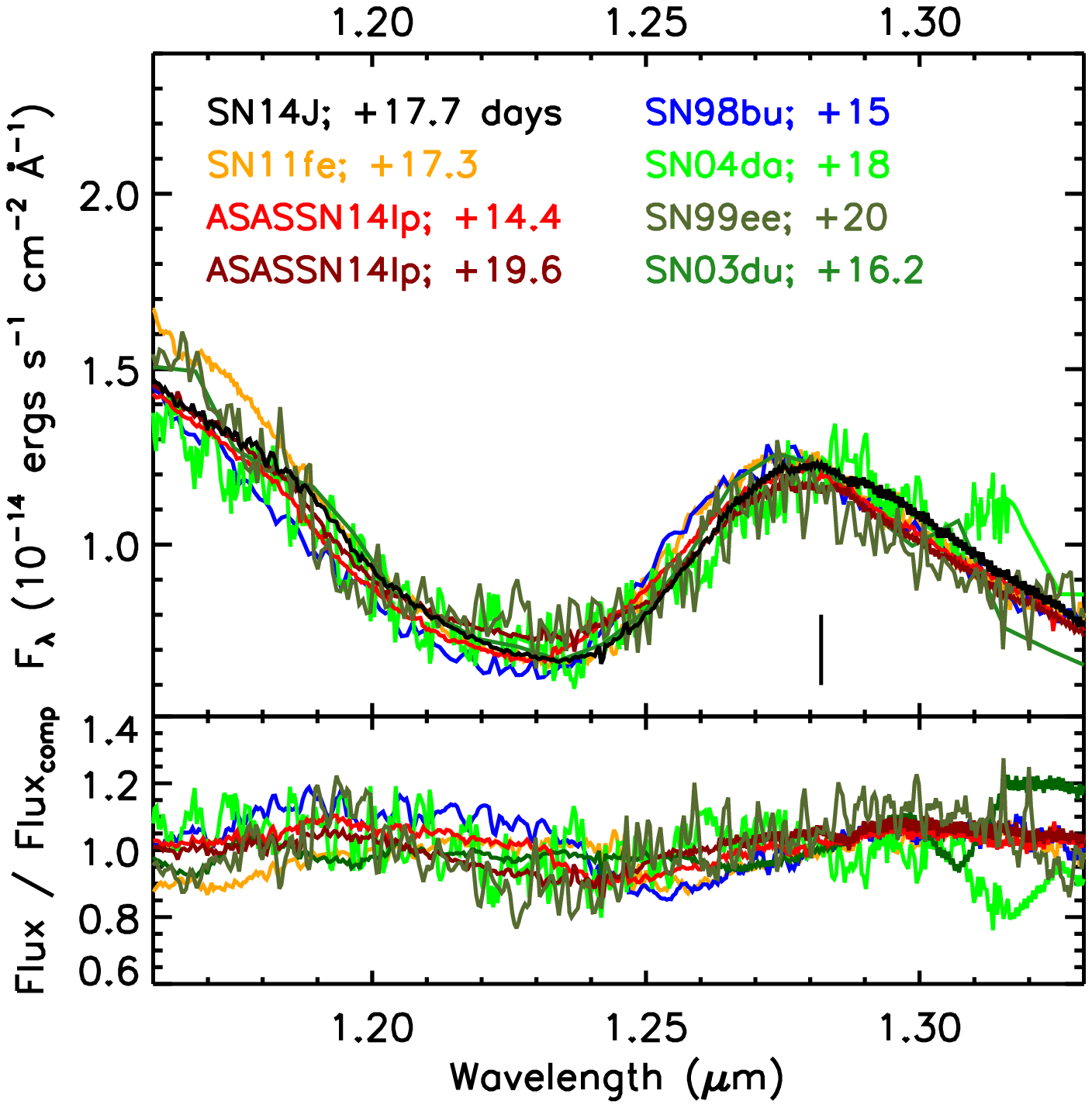}} 
\mbox{ \epsfysize=5.3cm \epsfbox{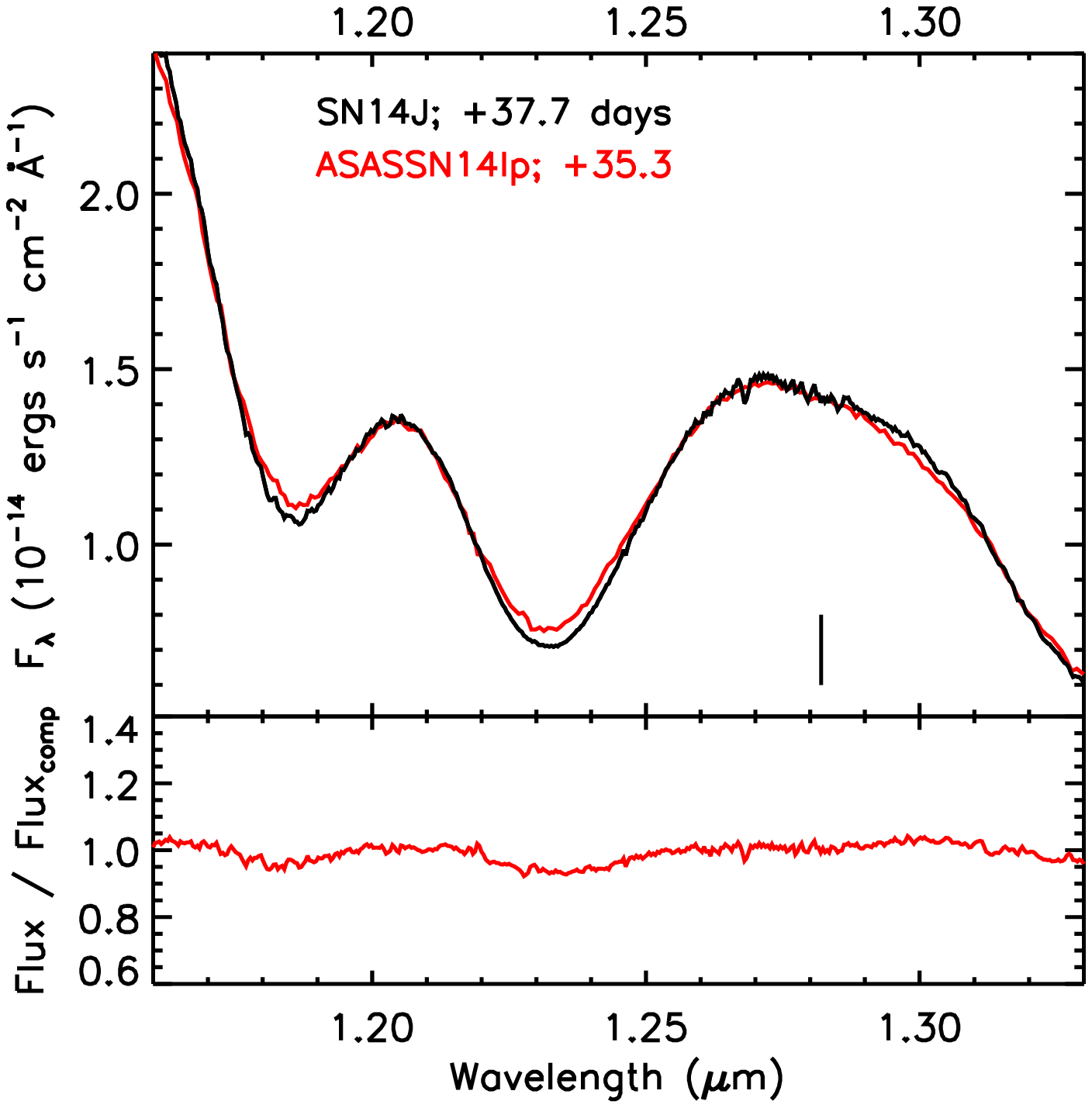}} 
\mbox{ \epsfysize=5.3cm \epsfbox{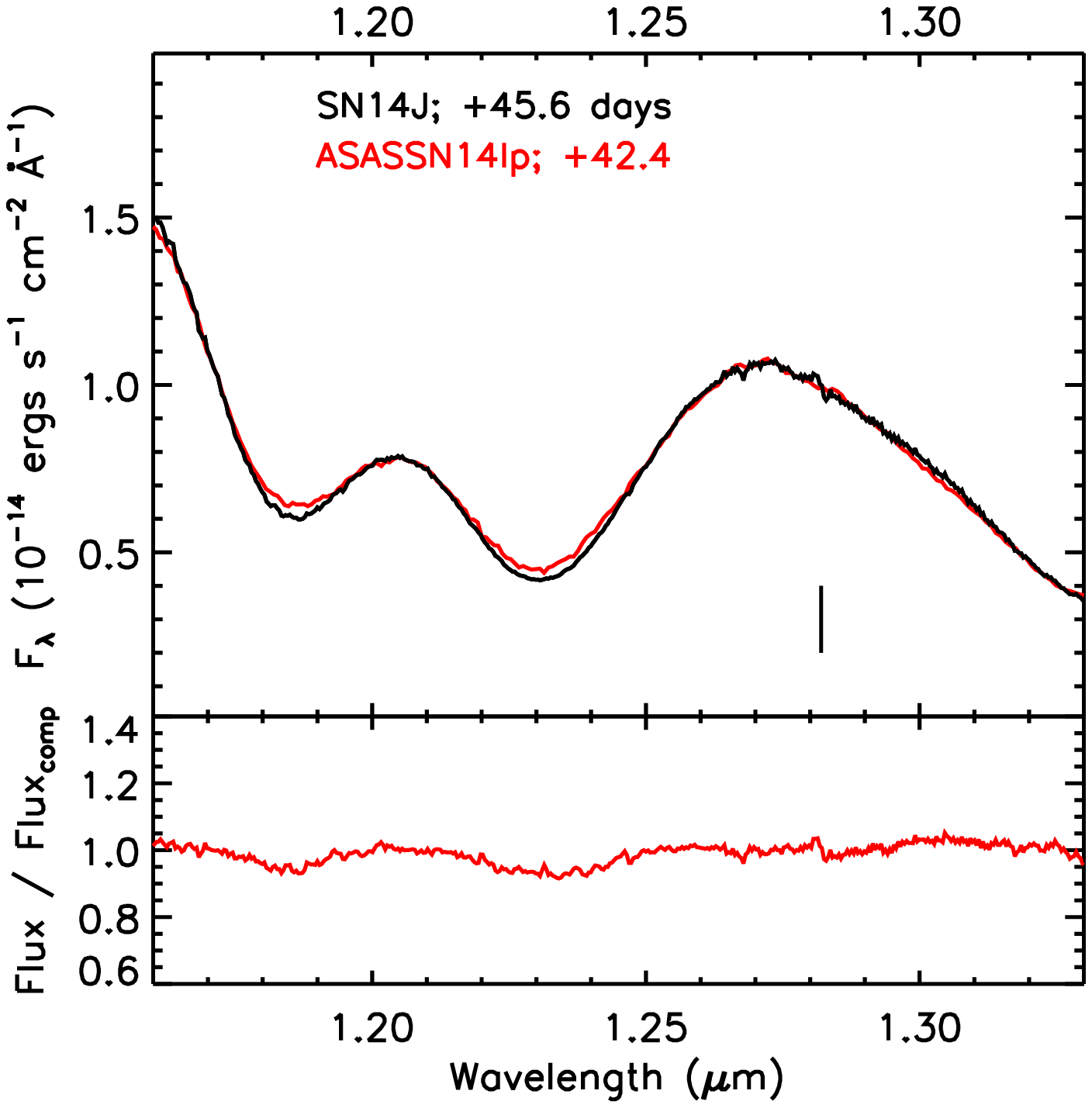}} 
\mbox{ \epsfysize=5.3cm \epsfbox{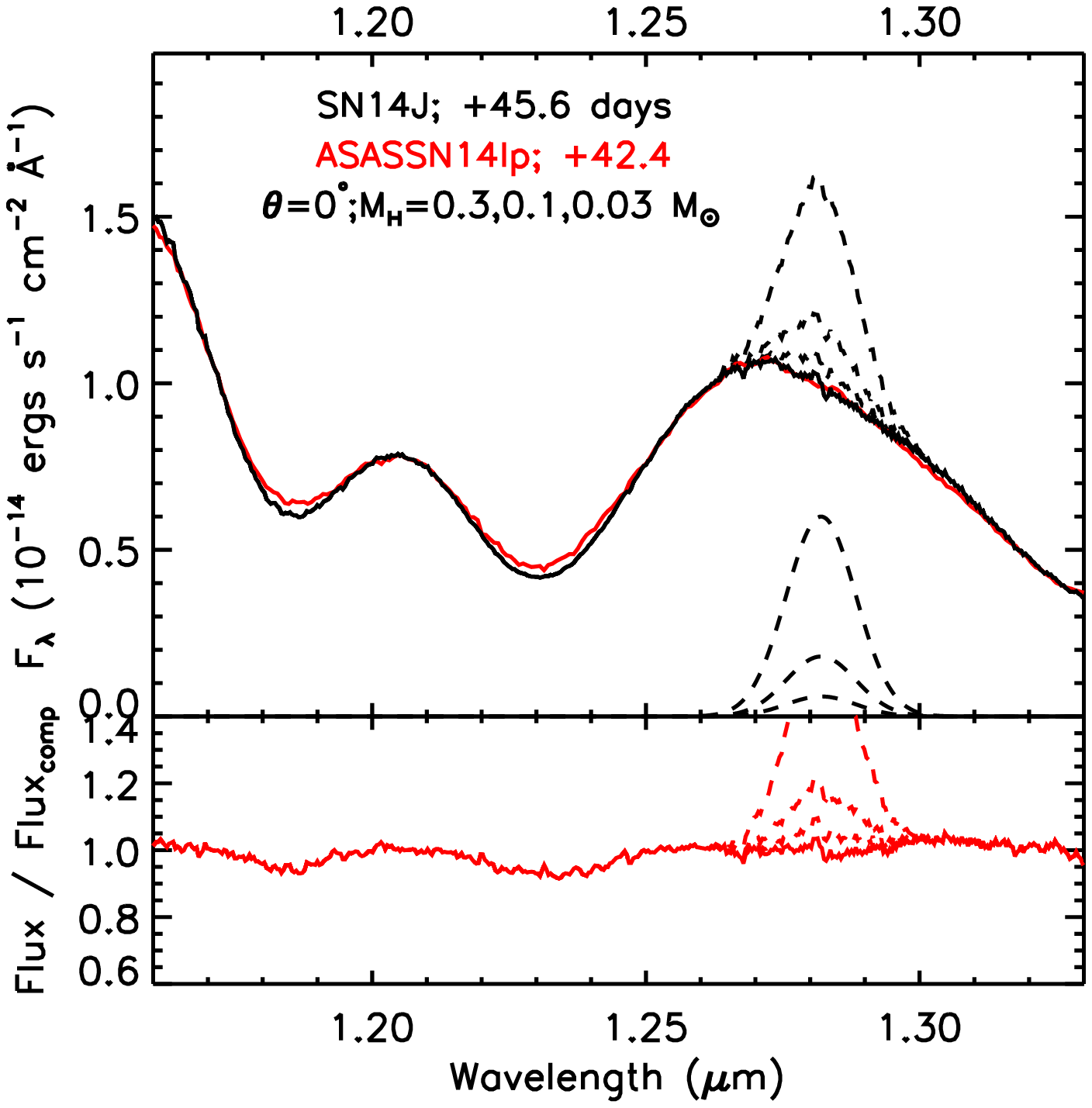}} 
\mbox{ \epsfysize=5.3cm \epsfbox{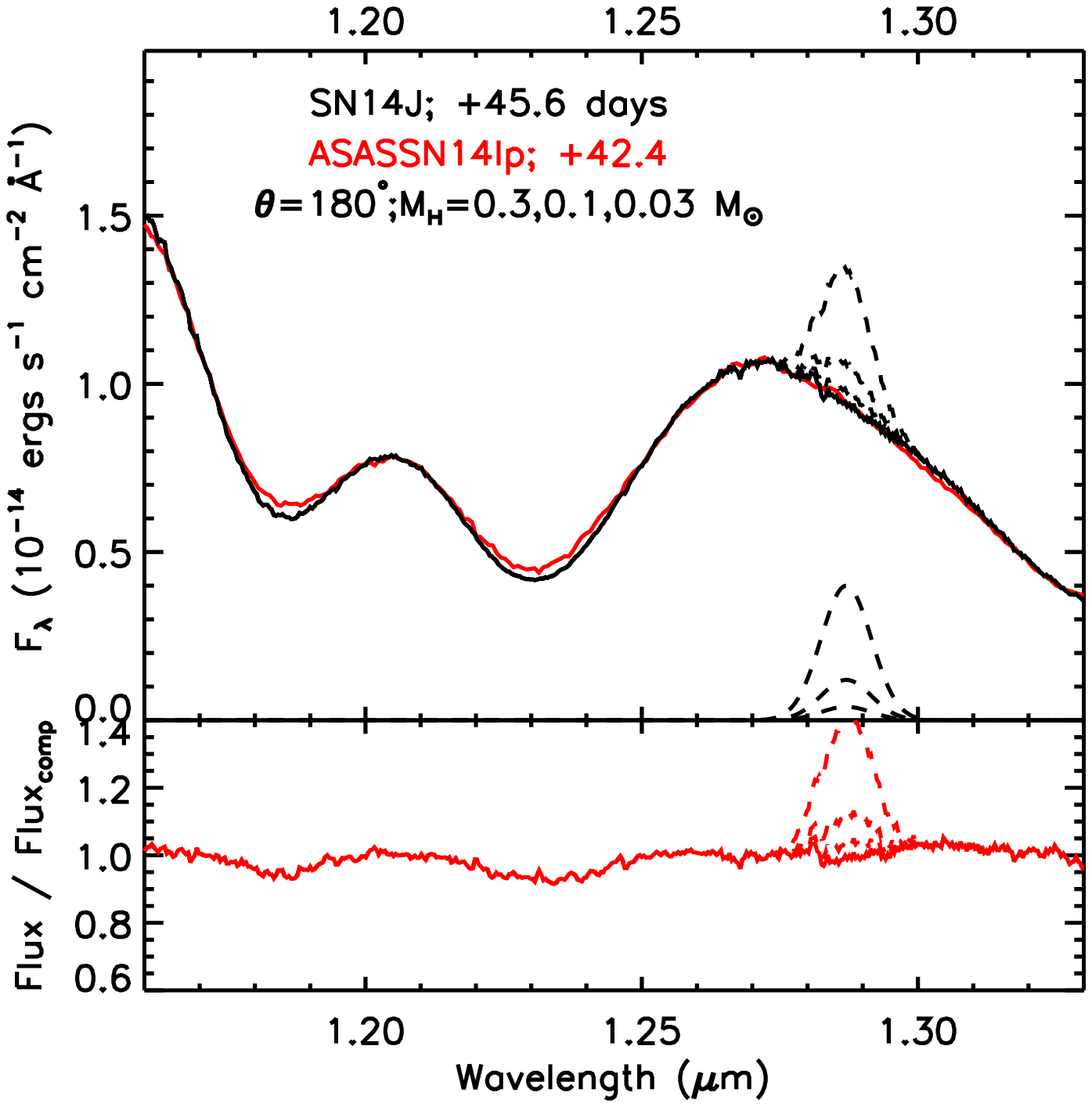}} 
\caption{{\bf Top Row} GNIRS spectra of SN~2014J in the region around Pa$\beta$, 1.282 $\mu$m, at +17.7d, +37.7d and +45.6d.  For each epoch, we match spectra in the literature with phases within $\sim$4 days of the SN~2014J epoch, and with $\Delta$$m_{15}$$(B)$$<$1.5 mag.  See \S~\ref{sec:search} for details.  In the bottom panel of each plot we divide the flux of the SN~2014J spectrum by the comparison spectrum, revealing no clear Pa$\beta$ emission at any epoch. {\bf Bottom Row} We have implanted the expected Pa$\beta$ emission as determined by M14 into our SN~2014J spectrum at +45.6d, for scenarios where the companion star is directly between the observer and the supernova ($\theta$=0$^{\circ}$) and directly behind ($\theta$=180$^{\circ}$).  In either case, unbound hydrogen masses down to $\sim$0.1 $M_{\odot}$ would be detectable in our data set. \label{fig:PB}}
\end{center}
\end{figure*}

For each phase-matched spectrum, we rescaled the flux to match that of SN~2014J 
in the 1.15 to 1.35 $\mu$m region and over-plotted the result (Figure~\ref{fig:PB}).  If we assume that each comparison spectrum serves as a `hydrogen-free' template, dividing the relevant SN~2014J spectrum by the template spectrum yields a residual spectrum that can be searched for features around the Pa$\beta$ line (also shown in Figure~\ref{fig:PB}).  It is not necessarily the case that each comparison spectrum is `hydrogen-free'.  If, for instance, the comparison spectrum had significant Pa$\beta$ emission but SN~2014J did not, then we would see an absorption feature in the residual spectrum (we do not see any such features).  If both the comparison spectrum and SN~2014J had similar amounts of Pa$\beta$ the residual spectrum would be smooth with no hydrogen sensitivity.  While we continue with the `hydrogen-free' conceit throughout the rest of the paper, this highlights the importance of obtaining larger samples of post-maximum NIR spectra of nearby SN Ia.

The comparison NIR spectra match that of SN~2014J reasonably well, and the ASASSN-14lp spectra in particular are nearly an identical match. 
A visual inspection of the residual spectra shows smooth variations, likely due to slight differences in the \ion{Fe}{2} line strength and velocity.  There are no emission features in the Pa$\beta$ region.


To put a quantitative limit on the mass of hydrogen in our SN~2014J spectra, we have implanted Gaussian emission lines into the Pa$\beta$ region corresponding to those predicted by M14 (in their red giant companion scenario, RGa).  To do this, we refer to Figure~23 in M14, where Pa$\beta$ emission lines drawn from their model spectra were implanted into the NIR spectrum of SN~1999ee at +39d (we did not use the SN~1999ee spectrum ourselves, as its morphology did not match that of SN~2014J).  We have visually determined the total flux and velocity width of the Pa$\beta$ lines in Figure~23 of M14, and rescaled them to the distance of SN~2014J.  There is potentially at least a $\sim$20\% systematic associated with this `by eye' estimation, even just factoring in the distance uncertainties of SN~1999ee and SN~2014J; this is acceptable for the purposes of this exploratory study.
For the scenario where the non-degenerate companion star is directly between the observer and the supernova ($\theta$=0$^{\circ}$), we estimate a hydrogen mass of $\sim$0.3 $M_{\odot}$ corresponds to a Pa$\beta$ flux of 6 $\times$ 10$^{-15}$ ergs s$^{-1}$ cm$^{-2}$ \AA$^{-1}$ and velocity full width at half maximum of 3500 km s$^{-1}$.  Likewise, for a viewing angle of $\theta$=180$^{\circ}$ a Pa$\beta$ flux of 4 $\times$ 10$^{-15}$ ergs s$^{-1}$ cm$^{-2}$ \AA$^{-1}$ and velocity full width at half maximum of 2500 km s$^{-1}$ was estimated from M14. While these values correspond to the optimistic red giant companion scenario of M14, we assume that changes in the Pa$\beta$ flux correspond to proportional changes in the hydrogen mass, as was implied by M14 themselves.

We focus on the +45.6d spectrum of SN~2014J, and its well-matched counterpart, ASASSN-14lp at +42.4d, as this phase is expected to have the strongest Pa$\beta$ emission over the phase range studied by M14.  In the bottom panels of Figure~\ref{fig:PB} we have implanted Pa$\beta$ emission lines corresponding to 0.3, 0.1, 0.03 $M_{\odot}$ of hydrogen for both $\theta$=0$^{\circ}$ and 180$^{\circ}$ scenarios.  Inspecting the residual spectra when the Pa$\beta$ lines have been implanted, it is clear $\sim$0.1 $M_{\odot}$ of unbound hydrogen can be ruled out for all SN-companion star orientations, and we adopt this as the hydrogen mass limit found in our study.  Hydrogen masses of  $\sim$0.03 $M_{\odot}$ are not distinguishable with our technique.




\section{Discussion \& Conclusions}  

We have presented an extensive NIR spectroscopic data set of the nearby SN~Ia 2014J.  A $JHK_s$ light curve is also presented that extends the dataset out to +71d, while the seventeen NIR spectra correspond to epochs between +15.3d and +92.5d.  We use the NIR spectra to search for Pa$\beta$ emission at these epochs to constrain any interaction with hydrogen rich material.

Visual inspection of the spectral region around Pa$\beta$, and direct comparisons with NIR spectra of other SNe Ia taken at similar epochs, yield no clear emission.  By placing Gaussian emission lines into our data, meant to mimic the red giant Pa$\beta$ models of M14, we estimate hydrogen mass limits of $\sim$0.1 $M_{\odot}$ for all viewing angles between the SN and any possible companion star (assuming a linear scaling between Pa$\beta$ flux and hydrogen mass).  These limits are comparable to or below the unbound mass seen in some simulations including a non-degenerate companion star \citep[e.g.][]{Liu12}, but higher than the unbound mass predicted and discussed by some others  \citep[e.g.][]{Pan12,Lundqvist15}.  As our limit is based off of the red giant model of M14, it is not clearly applicable to main sequence companions.    Thus, explicit modeling of Pa$\beta$ for a variety of red giant and main sequence companions would aid in the interpretation of any Pa$\beta$ found in the future.  Other effects, such as the velocity of the unbound hydrogen \citep[cf. ][]{Pan12}, can plausibly effect the line profile of Pa$\beta$ and should be modeled in detail as well.  


Other constraints on SN~2014J's progenitor system have lent tentative support to both the single \citep{Goobar15, Graham15} and double degenerate \citep{Kelly14,Margutti14,Perez14,Chomiuk15,Maeda16}  scenarios.  Most analogous to the current work, \citet{Lundqvist15} combined numerical models and a late time optical spectrum of SN~2014J (at +315d) to put a limit of 0.0085 $M_{\odot}$ on any hydrogen-rich unbound mass.  This limit is a factor of $\sim$10 more stringent than that presented here for the Pa$\beta$ line, but this should improve as both the spectral modeling and observational techniques become more sophisticated.  

The outlook for using post-maximum NIR spectra as a means to constrain the SN Ia progenitor is excellent.  In particular, the {\em Carnegie Supernova Project II} has acquired an extensive set of NIR spectra  on a variety of nearby SN Ia spanning their observational diversity \citep[e.g.][]{Hsiao15,Stritzinger15} from which a comprehensive analysis can be performed.  

\acknowledgments

DJS acknowledges support from NSF grants AST-1412504 and AST-1517649.  The research work at the PRL is funded by the Department of Space,  Government of India.  M. D. Stritzinger acknowledge support provided by the Danish Agency for Science and Technology and Innovation realized through a Sapere Aude  Level 2 grant.  This paper is partially based on observations carried out by the CSP that were supported by the National Science Foundation under Grant No. AST-1008343. We are grateful to Stefano Valenti for his comments.


\bibliographystyle{apj}


\clearpage
\LongTables

\begin{deluxetable}{rrrc}
\tablecolumns{4}
\tablecaption{NIR Photometry\label{table:nirphot}}
\tablehead{\colhead{MJD} & \colhead{Phase wrt} & \colhead{Mag} & \colhead{Err}  \\
                  \colhead{+56600} & \colhead{\bmax} & \colhead{Apparent} & \colhead{} }\\
\startdata
\cutinhead{$J$ band}
79.77 & -9.97 & 9.94 & 0.06 \\
80.83 & -8.91 & 9.70 & 0.03 \\
81.79 & -7.95 & 9.61 & 0.04 \\
82.76 & -6.98 & 9.53 & 0.03 \\
83.75 & -5.99 & 9.45 & 0.03 \\
84.71 & -5.03 & 9.45 & 0.07 \\
85.74 & -4.00 & 9.40 & 0.07 \\
86.76 & -2.98 & 9.41 & 0.03 \\
87.78 & -1.96 & 9.46 & 0.04 \\
90.78 & 1.04 & 9.63 & 0.04 \\
91.78 & 2.04 & 9.69 & 0.03 \\
92.79 & 3.05 & 9.92 & 0.03 \\
94.78 & 5.04 & 9.92 & 0.03 \\
95.78 & 6.04 & 10.06 & 0.02 \\
96.74 & 7.00 & 10.12 & 0.03 \\
97.74 & 8.00 & 10.34 & 0.03 \\
99.75 & 10.01 & 10.68 & 0.02 \\
100.75 & 11.01 & 10.91 & 0.04 \\
103.84 & 14.10 & 11.13 & 0.03 \\
104.86 & 15.12 & 11.15 & 0.03 \\
105.87 & 16.13 & 11.25 & 0.04 \\
106.78 & 17.04 & 11.18 & 0.03 \\
107.83 & 18.09 & 11.21 & 0.03 \\
109.75 & 20.01 & 11.15 & 0.02 \\
117.73 & 27.99 & 10.64 & 0.03 \\
119.73 & 29.99 & 10.57 & 0.03 \\
129.71 & 39.97 & 11.10 & 0.03 \\
130.72 & 40.98 & 11.27 & 0.03 \\
136.71 & 46.97 & 11.55 & 0.03 \\
141.71 & 51.97 & 12.05 & 0.04 \\
144.71 & 54.97 & 11.98 & 0.05 \\
146.73 & 56.99 & 12.20 & 0.05 \\
150.70 & 60.96 & 12.30 & 0.05 \\
151.70 & 61.96 & 12.51 & 0.05 \\
160.69 & 70.95 & 12.91 & 0.06 \\
\cutinhead{$H$ band}
79.76 & -9.98 & 9.83 & 0.06 \\
80.85 & -8.89 & 9.70 & 0.03 \\
81.82 & -7.92 & 9.67 & 0.04 \\
82.79 & -6.95 & 9.57 & 0.03 \\
83.76 & -5.98 & 9.57 & 0.04 \\
84.72 & -5.02 & 9.48 & 0.08 \\
85.74 & -4.00 & 9.54 & 0.07 \\
86.77 & -2.97 & 9.59 & 0.03 \\
87.80 & -1.94 & 9.58 & 0.04 \\
90.78 & 1.04 & 9.73 & 0.05 \\
91.78 & 2.04 & 9.75 & 0.05 \\
92.79 & 3.05 & 9.72 & 0.05 \\
94.80 & 5.06 & 9.68 & 0.05 \\
95.80 & 6.06 & 9.77 & 0.04 \\
96.75 & 7.01 & 9.77 & 0.04 \\
97.77 & 8.03 & 9.82 & 0.04 \\
99.77 & 10.03 & 9.80 & 0.05 \\
100.76 & 11.02 & 9.85 & 0.04 \\
103.86 & 14.12 & 9.79 & 0.04 \\
104.87 & 15.13 & 9.76 & 0.04 \\
105.88 & 16.14 & 9.82 & 0.05 \\
106.78 & 17.04 & 9.74 & 0.04 \\
107.85 & 18.11 & 9.71 & 0.03 \\
109.76 & 20.02 & 9.68 & 0.03 \\
117.74 & 28.00 & 9.52 & 0.04 \\
119.74 & 30.00 & 9.56 & 0.04 \\
129.72 & 39.98 & 10.05 & 0.04 \\
130.73 & 40.99 & 10.23 & 0.04 \\
136.72 & 46.98 & 10.30 & 0.04 \\
141.72 & 51.98 & 10.68 & 0.04 \\
144.72 & 54.98 & 10.50 & 0.05 \\
146.74 & 57.00 & 10.89 & 0.05 \\
150.71 & 60.97 & 10.70 & 0.04 \\
151.71 & 61.97 & 11.05 & 0.05 \\
160.70 & 70.96 & 11.45 & 0.04 \\
\cutinhead{$K_s$ band}
79.76 & -9.98 & 9.80 & 0.08 \\
80.88 & -8.86 & 9.53 & 0.06 \\
81.85 & -7.89 & 9.48 & 0.07 \\
82.82 & -6.92 & 9.40 & 0.05 \\
83.77 & -5.97 & 9.36 & 0.05 \\
84.72 & -5.02 & 9.27 & 0.10 \\
85.75 & -3.99 & 9.27 & 0.09 \\
86.77 & -2.97 & 9.25 & 0.04 \\
87.80 & -1.94 & 9.25 & 0.06 \\
91.78 & 2.04 & 9.35 & 0.07 \\
92.79 & 3.05 & 9.41 & 0.08 \\
94.81 & 5.07 & 9.41 & 0.07 \\
95.81 & 6.07 & 9.44 & 0.06 \\
96.77 & 7.03 & 9.45 & 0.06 \\
97.79 & 8.05 & 9.52 & 0.05 \\
99.79 & 10.05 & 9.48 & 0.06 \\
100.77 & 11.03 & 9.52 & 0.08 \\
103.87 & 14.13 & 9.56 & 0.07 \\
104.87 & 15.13 & 9.58 & 0.06 \\
105.88 & 16.14 & 9.54 & 0.06 \\
106.78 & 17.04 & 9.55 & 0.05 \\
107.85 & 18.11 & 9.58 & 0.05 \\
109.77 & 20.03 & 9.54 & 0.04 \\
117.75 & 28.01 & 9.50 & 0.04 \\
119.75 & 30.01 & 9.55 & 0.04 \\
129.73 & 39.99 & 10.18 & 0.04 \\
130.74 & 41.00 & 10.23 & 0.04 \\
136.73 & 46.99 & 10.35 & 0.04 \\
141.73 & 51.99 & 10.65 & 0.04 \\
144.73 & 54.99 & 10.65 & 0.04 \\
146.75 & 57.01 & 10.73 & 0.05 \\
150.72 & 60.98 & 10.98 & 0.05 \\
151.72 & 61.98 & 10.90 & 0.05 \\
160.72 & 70.98 & 11.35 & 0.05 \\
\enddata
\end{deluxetable}
\clearpage

\begin{deluxetable*}{ccrcrrccc}
\tablecolumns{9}
\tablecaption{Log of Spectroscopic Observations \label{table:spectlog}}
\tablehead{
 \colhead{UT Date} & \colhead{MJD} & \colhead{Phase wrt} & \colhead{Observatory/} &
 \colhead{N} & \colhead{I. Time} & \colhead{Airmass} & \colhead{Telluric/Flux} & \colhead {Airmass} \\
\colhead{} & \colhead{} & \colhead{\bmax\tablenotemark{a}} & \colhead{Instrument} & \colhead{ Exp } & \colhead{(s)} & \colhead{SN 2014J} & \colhead{Standard} & \colhead{Standard}
}
\startdata
2014-02-16 & 56705.02 & 15.28 & Abu & 4 & 720 & 1.42 & SAO14667 & 1.38 \\
2014-02-17 & 56706.02 & 16.28 & Abu & 4 & 760 & 1.42 & SAO14667 & 1.42 \\
2014-02-20 & 56708.35 & 18.61 & GNIRS & 16 & 960 & 1.65 & HIP52478 & 1.55 \\
2014-02-21 & 56710.06 & 20.32 & Abu & 8 & 1440 & 1.44 & SAO14667 & 1.46 \\
2014-03-01 & 56718.11 & 28.37 & Abu & 10 & 1800 & 1.56 & SAO14667 & 1.40 \\
2014-03-03 & 56720.06 & 30.32 & Abu & 10 & 1800 & 1.47 & SAO14667 & 1.35 \\
2014-03-11 & 56727.45 & 37.71 & GNIRS & 20 & 1200 & 1.62 & HIP52478 & 1.33\\
2014-03-13 & 56730.02 & 40.28 & Abu & 10 & 1500 & 1.45 & SAO14667 & 1.52 \\
2014-03-14 & 56731.02 & 41.28 & Abu & 10 & 1500 & 1.46 & SAO14667 & 1.41 \\
2014-03-15 & 56732.05 & 42.31 & Abu & 10 & 1500 & 1.52 & SAO14667 & 1.39 \\
2014-03-19 &56735.38 &45.64 & GNIRS & 28 & 1680 &1.56 &HIP52478 & 1.31\\
2014-03-19 & 56736.04 & 46.30 & Abu & 10 & 1800 & 1.51 & SAO14667 & 1.45 \\
2014-03-20 & 56737.05 & 47.31 & Abu & 10 & 1800 & 1.56 & SAO14667 & 1.49 \\
2014-04-03 &56750.28 &60.54 & GNIRS & 24 & 1440 &1.56 &HIP52478 &1.35\\
2014-04-23 &56770.25 &80.51 & GNIRS & 28 & 1680 &1.55 &HIP42434 &1.32\\
2014-04-23 & 56770.84 & 81.10 & Abu & 10 & 1800 & 1.42 & SAO14667 & 1.46\\
2014-05-05 &56782.24 &92.50 & GNIRS & 16 & 960 &1.55 &HIP52478 &1.27
\enddata
\tablenotetext{a}{MJD of $B_{max}$ = 56689.74 (Feb 01.74 UT).}
\end{deluxetable*}

\end{document}